\documentstyle[aps,twocolumn]{revtex}
\begin{document}
\draft

\begin{title}
{\bf Fluxon Pinning Through Interaction with the Superconducting Wiring
 of Long Annular Josephson Junctions}
\end{title}

\author{D. M\"unter,
T. Doderer \footnote{Present address:
IBM T. J. Watson Research Center, Yorktown Heights, New York, 10598, USA},
 H. Pre{\ss}ler \footnote{Present address:
Electrotechnical Laboratory, Umezono 1-1-4, Tsukuba, 305, Japan},
 S. Keil, and R. P. Huebener}
\address{
Physikalisches Institut,
Lehrstuhl Experimentalphysik II, Universit\"at T\"ubingen,
Auf der Morgenstelle 14, D-72076 T\"ubingen, Germany.
}

\date{\today}
\maketitle
\begin{abstract}
The statics and dynamics of magnetic flux quanta (fluxons) in long annular 
Josephson tunnel junctions have been investigated. Pinning by interaction 
of the fluxon field outside the junction with the superconducting wiring 
has been observed in spatially resolving 
measurements using low-temperature scanning electron microscopy. 
We were able to influence the characteristics of this field
by carefully modifying the beam-induced fluxon trapping procedure. 
In this way we were able to select the pinning site acting on the fluxon.

\vspace{25mm}

\end{abstract}

\section{Introduction}

The nonlinear system constituted by a Josephson tunnel junction has 
served as a rich source of interesting physical phenomena over the past
decades 
\cite{junctions}.
Especially long junctions, i.e. junctions with at least one dimension 
large against the Josephson penetration depth $\lambda_{j}$, have proven
to be a suitable tool for the investigation of soliton (fluxon) statics and 
dynamics. Most of the experimental work was carried out on linear Josephson 
junctions in the inline or overlap geometry
\cite{fluxon}.
The samples used for the experiments described in this article were of 
annular quasi-one-dimensional shape and are characterized in detail in 
section II.

One-dimensional long Josephson junctions are governed by the
perturbed sine-Gordon equation
\cite{mclaughlin}:
\begin{equation}
\varphi_{xx}-\varphi_{tt}=\sin{\varphi}+\alpha\varphi_{t}-\beta\varphi_{xxt}-\gamma+f(x,\varphi),
\label{PSGL}
\end {equation}
where $\varphi(x,t)$ is the spatially and time dependent phase difference
across the tunnel barrier between the two Cooper pair systems of the 
superconducting electrodes. In the case of annular Josephson junctions, $x$ 
runs azimuthally around the junction.
The variables $x$ and $t$ are normalized to the Josephson
penetration depth $\lambda_{j}$ and the inverse of the plasma frequency 
$\omega_{p}^{-1}$, respectively. The dissipation parameters $\alpha$ and 
$\beta$ account for losses due to quasi-particle currents across and parallel 
to the tunnel barrier, respectively, whereas $\gamma$ represents the normalized bias 
current. The function $f(x,\varphi)$ describes inhomogenities of various 
origins, such as externally applied magnetic fields or local changes of the 
thickness of the insulating barrier (e.g. microshorts). For annular junctions the dynamics and 
statics of solitons in the presence of inhomogenities have already been 
investigated experimentally and numerically. 
\cite {mclaughlin,keil,stephan,gronbech,martuciello1}. 
Because strictly speaking the term 'soliton' only refers to certain exact 
solutions of the {\em unperturbed} sine-Gordon equation, we use the word 
'fluxon' from this point on.
In the annular geometry the boundary conditions for Eq.(\ref{PSGL}) 
are periodic, i.e.
\begin{equation}
\varphi(l,t)=\varphi(0,t)+n2\pi\hspace{1cm}\varphi_{x}(l,t)=\varphi_{x}(0,t),
\label{Eindeutigkeit} 
\end {equation}
where $l$ is the ring circumference in units of $\lambda_{j}$ and $n$
is the net number of fluxons trapped in the junction. This means that there 
can only be an integer number of fluxons inside the junction at any given
time. If each of the two superconducting junction electrodes is 
thicker than the magnetic penetration depth $\lambda_{L}$ (which is the 
case for the samples studied here), the total magnetic flux and thus the net 
fluxon number in the annular junction is conserved. This conservation and 
the fact, that fluxons can move inside the junctions without influence of 
boundaries have motivated the experiments with annular junctions. 

Even though there have been investigations to determine the behavior of 
fluxons in externally applied magnetic fields, there have been no experiments so far, 
to our knowledge, considering the fluxon magnetic field outside the actual 
junction. In our experiments we found evidence
that an interaction of this field with the superconducting wiring takes
place and results in a pinning potential when junctions of the "Lyngby geometry"
\cite{annular1} 
are used, as is the case in most of the studies in the literature. We have 
also investigated the 
single-fluxon statics in this potential by means of low temperature 
scanning electron microscopy (LTSEM) which allows a spatially 
resolved measurement of the Josephson current. These results are 
presented in section III. Moreover, we discovered consequences of this
pinning in the dynamics of multifluxon systems to the degree that a 
consecutive transition of the fluxons into the dynamic state was observed 
as described in section IV. 

\section{The Samples and the Experimental Setup}

The samples that have been used for our measurements are
$\rm Nb/AlO_x/Nb$ tunnel junctions 
\cite{hypres} 
with the annular geometry introduced by the Lyngby group [see Fig. \ref{1}]
\cite{annular1}.
The current density $j_c$ at 4.2 K of these junctions is 1000 A/cm$^2$, 
the mean radius is 90 $\rm \mu m$. The width of the rings is 10 
$\rm \mu m$ (sample A) and 5 $\rm \mu m$ (sample B), respectively. 
The circumference of the ring, being the junction length, 
is 565 $\rm \mu m$ for both samples, corresponding to about 49 $\lambda_j$. 
This is very long compared to most other experiments conducted with annular 
junctions. This fact turns out to be vital for the observation of the phenomena 
described in this article. At 4.2 K the critical currents of the samples are approximately
45 mA (sample A) and 23 mA (sample B). The zero field step asymptotic voltages 
were measured to be about 30 $\rm \mu V$ at 4.2 K

We performed spatially resolved measurements by means of
low-temperature scanning electron microscopy (LTSEM)
\cite{ltsem1,ltsem2}.
LTSEM allows the local thermal perturbation of the junction due to the
electron beam with the focus at $x_0$ during operation of the sample at 
liquid helium temperatures. The spatial extension of this perturbation is 
given by the thermal healing length and its value determines the spatial 
resolution, being about 1--2\,$\rm \mu m$ for the samples studied here. 
The beam-induced temperature increment $\Delta T(x_0)$ can be tuned 
by the electron beam power.
Since the thermal relaxation time for the beam induced local thermal 
perturbation is about 100\,ns, the sample response signal is a time averaged 
information about the junction dynamics. The latter evolve on a time scale 
of about 10\,ps.

In this article we present two different imaging techniques for the pinned 
fluxons. The first method uses the electron beam induced additional energy 
loss during the collision of two fluxons. In case of fluxon-antifluxon collisions 
a significant electron beam induced voltage signal $\Delta V (x_0)<0$ of the 
current-biased junction at the collision sites can be observed 
\cite{laub,lachenmann}.
Notice, that we are dealing with the collision of fluxons of the same polarity 
in contrast to the fluxon-antifluxon collisions considered in Ref.
\cite{laub,lachenmann}.
Our experimental results show that also in the case of the collision of 
unipolar fluxons, the local thermal perturbation due to the electron beam 
causes a significant sample response $\Delta V(x_0)<0$ at the collision 
sites, as it has been already observed by Keil {\em et al}.
\cite{keil}

The second method is based on the two-dimensional imaging
of the spatial distribution of the maximum dc Josephson current density
\cite{bosch,quenter}
\begin{equation}
j(x)=j_c(x) \sin \varphi (x)
\label{J}
\end{equation}
depending on $j_c$ and on the phase difference $\varphi$. In our case, 
where $j_c$ does not show any significant spatial dependence across the 
junction
\cite{jc}, 
$j(x)$ is directly proportional to $\sin \varphi (x)$. The local 
thermal perturbation induced by the electron beam at $x_0$ decreases 
$j_c(x_0)$ and, neglecting any nonlocal effect, the change $\Delta j(x_0)$
is a direct measure of $\sin \varphi (x_0)$
\cite{bosch}.
The junction is biased close to the total critical current
\begin{equation}
I_c \approx I = \int_0^{l} j_c \sin \varphi(x) \, dx
\label{Ic}
\end{equation}
and $I_c$ is continuously measured during scanning the e-beam across 
the sample. We obtain the electron-beam-induced signal
\begin{equation}
- \Delta I_c(x_0) \propto \sin \varphi (x_0),
\label{DeltaIc}
\end{equation}
if the area perturbed by the beam is small compared to $\lambda_j$.
In case of a hysteretic current-voltage-characteristic ($IVC$) this imaging 
technique is described in 
\cite{bosch},
whereas for a nonhysteretic $IVC$ a description of the imaging technique 
can be found in \cite{quenter}.
For all images shown in this article any nonlocal effect of the local 
perturbation due to the macroscopic quantum properties of a Josephson 
junction can be ruled out
\cite{nonlocal}.
For a comprehensive review of the imaging of Josephson junction 
dynamics see Ref.
\cite{Thomas}.

\section{Single-Fluxon Experiments}

The key prerequisite for studying fluxon motion in annular Josephson 
junctions is fluxon trapping. For this purpose one has to break superconductivity. 
In addition, a magnetic field has to be applied during cooling of the sample 
through $T_c$. One or more fluxons can be trapped during this procedure. 
There are several methods for fluxon trapping 
\cite{ust1,ver}.
Since in a scanning electron microscope we investigate the cold samples, 
a reliable way of introducing fluxons into the junction consists of locally 
heating the superconducting electrodes to a temperature above $T_c$ 
by electron beam irradiation. A magnetic field can be applied during 
cooling by passing some current through the junction or by means of an 
external solenoid. In our experiments we found it favorable to apply an 
external field to achieve controllable trapping of single fluxons. Figure \ref{1} 
shows schematically the procedure of introducing a fluxon into the junction. 
After the scanning unit of the LTSEM is adjusted to deflect the electrons 
onto one of the electrodes the beam is switched on. This operation guarantees 
that the area heated above $T_c$ by the beam does not contain any magnetic 
flux because it is generated well apart from the edges in the superconducting 
material. Subsequently, when the electron beam is moved towards the 
center of the ring, the second electrode is heated as well. In this electrode 
magnetic flux supplied by the external field can be dragged along. Note 
that the different starting points depicted in Fig. \ref{1} (a) and (b) lead to 
different configurations of the external fluxon field. Inside the junction 
the results of the two procedures are identical.

After fluxon trapping, the number of trapped fluxons can easily be determined 
experimentally by recording the $IVC$ of the junction through applying a 
current to the junction. Due to the Lorentz force between the bias current and 
the magnetic moment of the fluxon the latter is accelerated, moves through 
the junction, and a voltage drop is observed. The $IVC$ of sample A
with one fluxon trapped by the procedure described above is shown in
Fig. \ref{2} (a) as well as the differential resistance ($dU/dI$) measured through
electronic differentiation with a lock-in amplifier. The asymptotic 
voltage of about 30 $\rm \mu V$ corresponds to the limiting Swihart 
velocity $\bar{c}$
\cite{swihart}, 
the velocity of light in the junction transmission line. 
Note that the curvature of the $IVC$ does not comply with perturbation 
theory, which predicts $\rm{d^{2}U}/{dI^{2}}<1$ for all V. We believe
that the reason for this is the voltage dependency of the quasiparticle tunneling 
probability 
\cite{tinkham},
which means that the dissipation parameter $\alpha$ is voltage dependent 
as well. 

The small critical current of about 1 mA (roughly 3 $\%$ of $I_c$ 
without fluxons trapped) suggests the existence of a pinning potential. 
Various authors have reported similar observations
\cite{annular1,ust2,martuciello2,harald}. 
Davidson {\em et al.} found a small critical current for the $IVC$ of a 
junction with a length of 15 $\lambda_j$ when fluxons were trapped.
That critical current depended on the junction history 
\cite{annular1}.

The statics of a single fluxon in the pinning potential responsible for the 
small critical current described above was investigated by LTSEM. In
Fig. \ref{3} (b) and (c) LTSEM images of a single fluxon trapped by the 
procedure depicted in Fig. \ref{1} (a) are shown (sample A). They show a 
single fluxon at two different positions in the ring. The white areas correspond 
to a Josephson supercurrent into the paper plane, the black areas into the opposite 
direction.
The magnetic moment and the direction
in which the Lorentz force acts are symbolized by arrows. Apparently the 
location of the pinning center depends on the direction of the bias current.
Considering that, because of the way the fluxon was trapped, the magnetic field 
of the fluxon encloses only the upper electrode (see Fig. \ref{1} (a)).Therefore, we
conclude that the fluxon gets pinned through the interaction with the upper electrode. 

Further evidence for the validity of this argument is given
by the LTSEM images of Fig. \ref{4}. The fluxon shown in these images  was 
trapped in the junction using the procedure depicted in Fig. \ref{1} (b) (sample A). Hence 
the magnetic field of the fluxon encloses only the lower electrode and 
gets caught on that wiring when acted on by the bias current. The fact, 
that the position of the pinned fluxons with respect to the horizontal symmetry 
axis of the junction
differs for reverse bias current directions in both Figures \ref{3} and \ref{4}, hints 
at the existence of additional forces on the fluxon. The measurements took 
place in a reasonably well shielded environment. However it is possible, that 
some external magnetic field is present at the sample site, which would 
act as an additional pinning force. 

An investigation of single fluxons trapped in the junction by crossing 
both electrodes {\em from the side}, i.e. switching on the electron beam outside 
the electrodes and then moving it into the center of the ring by crossing the narrowest
part of the combined electrodes, confirms the above interpretation. 
This procedure allows the field of the fluxon to circle both electrodes 
as schematically shown in Fig. \ref{5}. 
Even though shielding currents flow in the electrodes to ensure the 
unambiguity of the Cooper pair function in each electrode (the magnetic 
flux contained by a single fluxon is exactly the elementary flux quantum $\Phi_0$),
pinning is still expected to occur. The reason is the fluxon field outside the junction 
in the shape of a distorted magnetic dipole field having its maximum value just at the outer
border of the junction (see Fig. \ref{5} (b)). The distortion is due to the Meissner 
effect in the superconducting electrodes. Inside the junction the fluxon 
field is localized to an azimuthal length comparable to the Josephson penetration depth.
In this case interaction of the fluxon magnetic field with the wiring of both electrodes takes place. 
The resulting pinning potential is expected to be shallower than the potential discussed 
earlier for the procedures in Fig. \ref{1} (a) and (b). Consequently the critical current 
should be lower. Experiments showed indeed, that trapping of single fluxons over 
the sides resulted in a value of $I_c$ being 50 $\%$ smaller than for trapping from the 
middle of the electrodes.

It is interesting that the transition into the dynamic state does not happen in an 
abrupt manner. Even if a significant voltage drop is measured across the junction for a
single-fluxon mode (for example a voltage greater than 5 $\rm \mu V$ in Fig. \ref{2} (a)) 
the LTSEM images (not shown here) still display the Josephson current distribution 
of a fluxon at rest. We believe that this is a result of the relatively weak pinning 
forces which allow the fluxon to get depinned only part of the time at first. As described
in section II the LTSEM images are time averaged measurements. Therefore, the 
images recorded at finite voltages can be interpreted as time averaged images of a 
sporadically moving fluxon.

\section{Multifluxon Experiments}

The features of $IVCs$ do not change qualitatively when more than one fluxon 
is trapped. Figure \ref{2} (b) shows the $IVC$ and the differential resistance 
for the case of two fluxons trapped by the procedure shown in Fig \ref{1} (a). The 
$dU/dI$-curve displays two peaks at low voltages corresponding to the consecutive 
transition of the two fluxons from the static into the dynamic state. Note that $I_c$ 
of the two fluxon system is only about 60 \% of the value for one fluxon. It 
appears that this is due to the interfluxon repulsion of the fluxon 
magnetic moments. 
\cite{keil}. 

In the light of the concentration of the extrajunctional fluxon field in the 
immediate vicinity of the junction (see Fig. \ref{5} (b)) it is noteworthy that it 
was not possible to trap more than seven fluxons in the junction by repeatedly 
applying the procedures shown in Fig \ref{1}. 
On the other hand it needed only little effort to trap up to 50 fluxons by introducing 
them over the sides i.e. allowing the fluxon magnetic field to enclose both electrodes
\cite{harald}.
As already explained in section III (reduced) pinning is still expected for this case. 
The $IVC$ and the $dU/dI$-curve for such a multifluxon mode show a rather 
complicated shape as depicted in Fig. \ref{6} for the situation of seventeen fluxons 
trapped in the junction (sample B). Due to the presence of a small ohmic resistance in series
with the junction in this particular measurement the $IVC$ is tilted and the asymptotic voltage 
is larger than the 510 $\rm \mu V$ expected for a seventeen fluxon system. Nevertheless the 
structure of the $dU/dI$-curve can be explained by the consecutive transition of 
all seventeen fluxons from the static into the dynamic state.

Fluxon-fluxon collisions are expected for the situation where more than one fluxon is 
trapped in the junction, while at least one of them is moving and the others 
are pinned. A large number of LTSEM images were recorded near the resonant structures in 
Fig. \ref{6}. Arrows indicate the voltages at which selected images, shown in Fig. \ref{7}, were taken. In the LTSEM images we observe the following process.
One after another the fluxons get depinned when the bias current is increased, as indicated
by the decreasing number of signal peaks. Finally 
only a single one is still at rest while sixteen are moving as seen in Fig. \ref{7} (d).
The moving fluxons are used as a detector for the pinned ones by the 
collisions that take place once during each revolution of every moving fluxon. 

At this point it is necessary to recall that the transition into the dynamic state is not 
abrupt. Instead a sporadic change between statics and dynamics of a particular fluxon 
takes place. Therefore it makes sense that the image in Fig.\ref{7} (d) still shows the last 
fluxon, even though it was recorded at a voltage greater than the one corresponding 
to the last peak in the differential resistance in Fig. \ref{6}. This is in accordance with 
the findings for single trapped fluxons.
It is interesting that the position of the last static fluxon shown in Fig. 
\ref{7} (d) is identical with the one of a single fluxon getting pinned at the wiring of 
the upper electrode. For reversed bias current direction 
the pinning of the last fluxon at the other side of the same wiring was observed.

Furthermore we want to point out the way in which the size of the fluxons is different 
in each of the images of Fig. \ref{7}. The fluxon compression (see eq. \ref{Eindeutigkeit}
results in a reduced fluxon length, since the fluxons are 'squeezed' into the junction. 
The more fluxons change to the dynamic state, the less compression the static ones experience. 
Therefore the static fluxons expand. Finally the last static fluxon (Fig. \ref{7} (d)) reaches
the 'relaxed' size, i.e.the size we observe if only a single fluxon exists in the junction at rest,
being approximately 2$\pi \lambda_j$.

\section{Conclusions}

We have investigated the statics and dynamics of (Josephson) fluxons in long 
annular $\rm Nb/AlO_x/Nb$ tunnel junctions. The current-voltage characteristics 
showed evidence of fluxon pinning. Applying low temperature scanning electron 
microscopy, we were able to image the statics of single pinned fluxons. 
By carefully controlling the procedure through which fluxons were introduced 
into the junction, it was possible to influence the position of the pinned fluxons.
These experiments show that this pinning is a result of the interaction of the 
magnetic fluxon field with the superconducting wiring of the
junction. We have also investigated the behavior of multifluxon states. For 
the case of 17 fluxons trapped in the junction we found that the fluxons get 
depinned one after another with increasing bias current.

\section*{Acknowledgments}

We thank A. Laub for helpful discussions and G. M. Fischer for presenting 
us with a circuit fabricated by HYPRES, Inc.
Financial support from the Forschungsschwerpunktprogramm des Landes
Baden-W\"urttemberg is gratefully acknowledged.

\begin{figure}
\caption{Schematic representation of the trapping procedures for fluxons 
resulting in a defined extrajunctional magnetic field configuration. The 
electron beam is switched on when already focused on one of the electrodes. 
The energy deposited by the electron beam results in local heating to
temperatures above $T_c$. Magnetic flux can only penetrate the normal
region in the electrode crossed second by the beam. (a) shows the procedure 
starting from the bottom electrode, (b) from the top electrode.}
\label{1}
\end{figure}

\begin{figure}
\caption{Current-voltage curve ($IVC$) and differential resistance ($dU/dI$)
 of the annular junction (sample A) with (a) one fluxon and (b) two fluxons trapped using 
the trapping procedure of Fig. 1 (a). Small critical 
currents of about 1 mA and 0.6 mA, respectively, (approx. 3 $\%$ of $I_c$ with no 
fluxon) and the peaks in the $dU/dI$ curves are evidence of fluxon pinning. $T\approx 4.2$ K.}
\label{2}
\end{figure}

\begin{figure}
\caption{(a) Sketch of the annular junction with the same orientation as 
in (b) and (c) but drawn to different scale. The voltage images in (b) and 
(c) show a single static fluxon trapped by the procedure of Fig 1 (a) (sample A). 
For imaging, the junction was biased at I = +1.9 mA and I = -2.2 mA, 
respectively. Positive bias current values denote a current flow from the 
top to the bottom electrode. Fluxon magnetic moment and the direction 
of the Lorentz force are symbolized by arrows.}
\label{3}
\end{figure}

\begin{figure}
\caption{LTSEM voltage images of a single static fluxon trapped by the procedure 
depicted in Fig. 1 (b). The junction (sample A) was biased at I = +1.7 mA (a) and 
I = -1.6 mA (b), respectively. In contrast to the single-fluxon configuration 
shown in Fig. 3 pinning at the edges of the lower electrodes can be 
observed (sample orientation as shown in Fig. 3 (a)).}
\label{4}
\end{figure}

\begin{figure}
\caption{(a) Schematic drawing of a possible extrajunctional field configuration
for the case of a fluxon trapped by crossing both electrodes simultaneously with the
focused electron beam at their narrowest point, i.e. along the vertical symmetry axis 
in Fig. 3 (a). Shielding currents in the electrodes are necessary 
to ensure the unambiguity of the Cooper pair wave functions, which results in flux
quantization in the rings of the electrodes. (b) Sketch 
of the flux distribution for the upper electrode in the situation depicted 
in (a). Even though the net flux in the opening $\Phi_{net}=0$, the
fluxon magnetic field outside the junction will still interact with both wirings.}
\label{5}
\end{figure}

\begin{figure}
\caption{Current-voltage characteristic ($IVC$) and differential resistance (dU/dI)
of the junction (sample B) with 17 fluxons trapped. Due to a small ohmic 
resistance in series with the junction the$IVC$ is tilted. The peaks in the dU/dI-curve 
are a result of the consecutive transition of the fluxons from the static to the dynamic state. 
The arrows indicate the voltages at which the LTSEM images in Fig. 7 were recorded.}
\label{6}
\end{figure}

\begin{figure}
\caption{LTSEM voltage images of 17 fluxons trapped in the junction (sample B) 
at different bias current values. (sample orientation as shown in Fig. 3 (a)) 
(a) All 17 fluxons are at rest. The size of the fluxons
is reduced compared to the single-fluxon case of Fig. 3 and 4 due to fluxon compression
inside the junction. (b) At I = 2.1 mA the image still shows eleven signal peaks, indicating that 
six fluxons are moving. (c) Four peaks are left at a bias current I = 3.6 mA. (d).
Finally at I = 7.3 mA only a single static fluxon is visible.}
\label{7}
\end{figure}


\begin{references}
\bibitem{hypres}
The samples designed by the authors have been fabricated by HYPRES 
Inc., Elmsford, NY 10523.
\bibitem{jc} This can be checked using LTSEM by measuring the spatial distribution 
of the quasiparticle tunneling current density across the junction
\end{references}
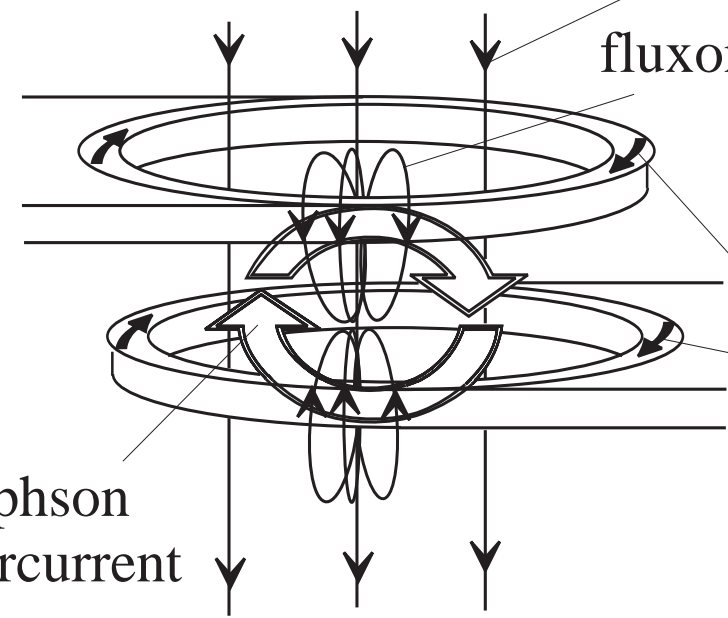
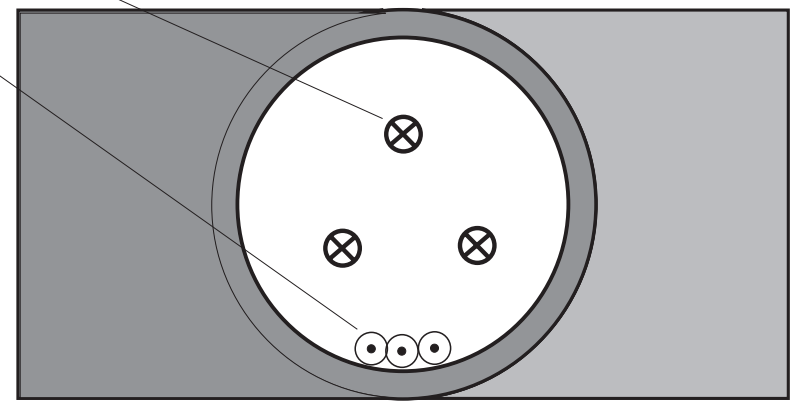
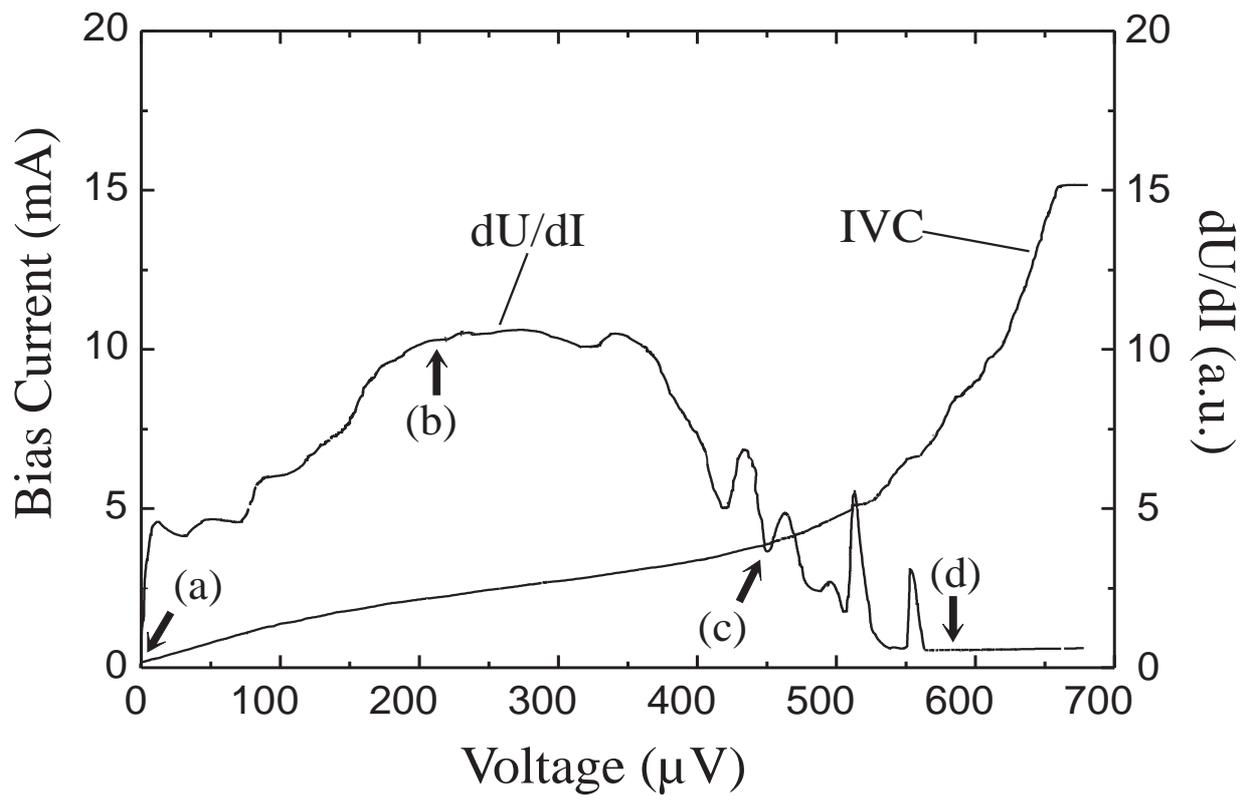
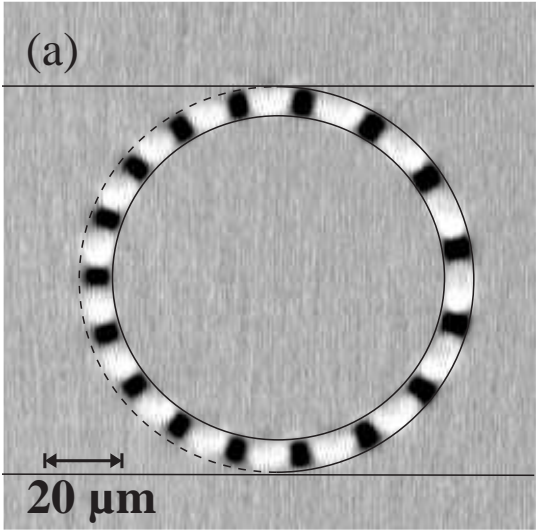
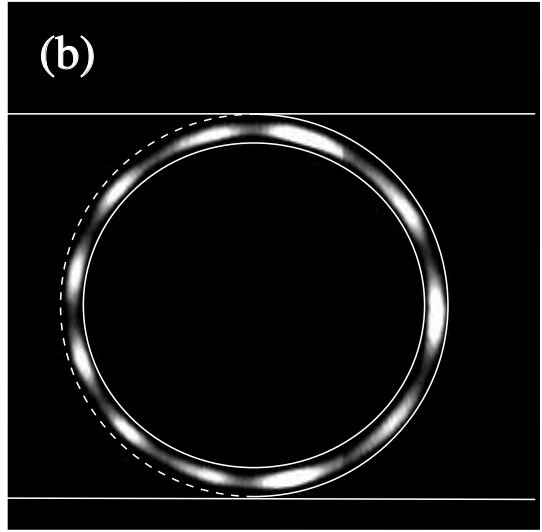
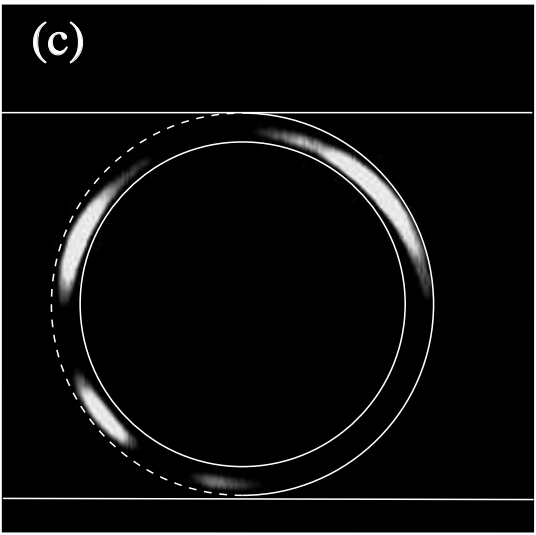
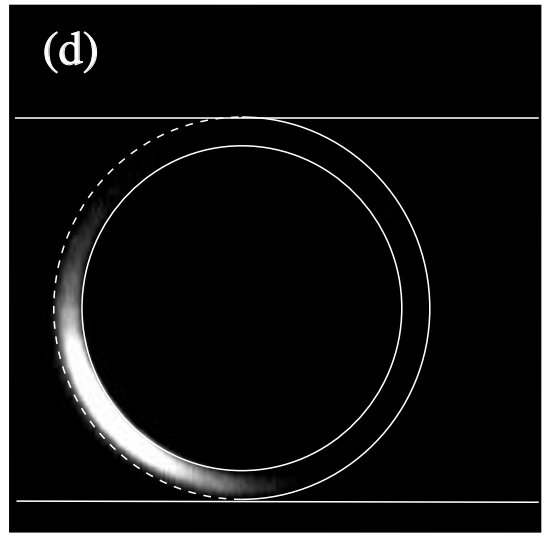
\end{document}